# Toward Sustainable Generative AI: A Scoping Review of Carbon Footprint and Environmental Impacts Across Training and Inference Stages

Kim Min-Kyu[1], Yoo Tae-An[2], Chung Ji-Bum[1,2, *]

[1] Department of Civil, Urban, Earth, and Environmental Engineering, Ulsan National Institute of Science and Technology (UNIST), Ulsan, Republic of Korea

[2] Graduate School of Carbon Neutrality, Ulsan National Institute of Science and Technology (UNIST), Ulsan, Republic of Korea

[*] Corresponding author: Ji-Bum Chung (learning@unist.ac.kr)

**Abstract**

Generative AI is spreading rapidly, creating significant social and economic value while also raising concerns about its high energy use and environmental sustainability. While prior studies have predominantly focused on the energy-intensive nature of the training phase, the cumulative environmental footprint generated during large-scale service operations, particularly in the inference phase, has received comparatively less attention. To bridge this gap this study conducts a scoping review of methodologies and research trends in AI carbon footprint assessment. We analyze the classification and standardization status of existing AI carbon measurement tools and methodologies, and comparatively examine the environmental impacts arising from both training and inference stages. In addition, we identify how multidimensional factors such as model size, prompt complexity, serving environments, and system boundary definitions shape the resulting carbon footprint. Our review reveals critical limitations in current AI carbon accounting practices, including methodological inconsistencies, technology-specific biases, and insufficient attention to end-to-end system perspectives. Building on these insights, we propose future research and governance directions: (1) establishing standardized and transparent universal measurement protocols, (2) designing dynamic evaluation frameworks that incorporate user behavior, (3) developing life-cycle monitoring systems that encompass embodied emissions, and (4) advancing multidimensional sustainability assessment framework that balance model performance with environmental efficiency. This paper provides a foundation for interdisciplinary dialogue aimed at building a sustainable AI ecosystem and offers a baseline guideline for researchers seeking to understand the environmental implications of AI across technical, social, and operational dimensions.

# 1. Introduction

Artificial intelligence (AI) has rapidly expanded across diverse domains, ranging from machine learning and deep learning to the recent wave of generative AI, and has generated substantial social and economic value. At the same time, concerns over energy consumption and greenhouse gas (GHG) have intensified, fueling ongoing debates about whether AI serves as a net positive or a net negative for the environment (Gaur et al., 2023; S. Luccioni et al., 2025a; Rasheed et al., 2024; Wang et al., 2024). Proponents of the former argue that AI-driven optimization can reduce energy and resource use, whereas critics contend that the expansion of AI technologies creates new demand and infrastructure growth that ultimately increases the overall environmental burden. In this context, accurately assessing the environmental impacts of AI systems to inform decision making has become a critical challenge for both academia and industry.

Recent studies project a sharp increase in energy demand driven by the expansion of AI. In the United States, data center electricity consumption rose from 1.9 percent of total national usage in 2018 to 4.4 percent in 2023, and it is expected to reach between 6.7 and 12.7 percent by 2028 (Shehabi et al., 2024). More than half of this consumption is anticipated to be attributable to AI, and AI alone is projected to require electricity equivalent to 22 percent of all US households (Shehabi et al., 2024). Although the absolute scale remains smaller than that of electric vehicles or building electrification, the growth rate and geographic concentration of AI-related demand may impose substantial burdens on power grids, electricity prices, and environmental systems. In particular, demand for inference is far greater than that for training and may account for more than 90 percent of the entire computing cycle for large language models (LLMs), indicating that inference may emerge as the dominant long-term driver of electricity consumption rather than training (O'Donnell & Crownhart, 2025; Patel et al., 2024). However, because companies do not systematically disclose energy data, considerable uncertainty remains in estimating actual electricity usage and associated carbon emissions.

Environmental impact assessments are generally approached along two major axes: carbon footprint analysis and life cycle assessment (LCA). Carbon footprint analysis relies on standardized methodologies such as ISO 14067 and the GHG Protocol, allowing climate change impacts to be estimated quickly and consistently in terms of carbon emissions. This approach offers strong advantages for comparison and communication. In contrast, LCA evaluates a broader set of environmental indicators, including carbon, water use, toxicity, and land use, across the entire life cycle of a product or system, from raw material extraction to production, use, and end of life, often described as a cradle-to-grave perspective (Berthelot et al., 2024; Kurisu, 2015). Because of its broader scope, LCA requires more extensive data and involves greater complexity. Its results are highly sensitive to boundary definitions and underlying assumptions. In practice, for integrated services such as generative AI, it is common to begin by estimating the more straightforward carbon footprint and then extending the analysis to a full life cycle perspective when needed.

Carbon emissions are commonly classified along two complementary perspectives. The first perspective is based on responsibility and supply chain boundaries, following the GHG Protocol's Scope 1, Scope 2, and Scope 3 categories. Scope 1 covers direct emissions from facilities and equipment owned or controlled by a company. Scope 2 refers to indirect emissions associated with purchased electricity and heat. Scope 3 includes all other indirect emissions across the upstream and downstream supply chain, such as component manufacturing, logistics, product use, and end of life. For AI systems, emissions from model training and inference occurring within a company's own data centers are typically classified as Scope 1 or Scope 2, whereas emissions associated with cloud outsourcing are categorized as Scope 3 from the company's perspective. The second perspective classifies emissions by timing and activity, distinguishing between embodied emissions and operational emissions (ITU, 2025). Embodied emissions refer to emissions generated during the manufacturing, construction, deployment, and end-of-life stages of GPUs, TPUs, servers, and data center infrastructure. Operational emissions arise repeatedly during service operation, such as electricity consumption and cooling, and include the emissions generated during the training and inference stages of AI models (see Figure 1). Many studies argue that for AI systems, operational emissions constitute the dominant share of total carbon emissions compared with embodied emissions (Liu & Zhai, 2025; Schneider et al., 2025).

Existing research has traditionally focused on the intensive energy use of the training phase and the energy demands associated with developing large-scale models. However, as generative AI becomes widely adopted and user interactions surge, cumulative energy consumption in the inference phase is rapidly emerging as a major issue. In earlier machine learning applications, inference demand was relatively limited, but in today's environment, where LLMs process billions of requests per day, the usage phase has become as significant as, or even more significant than, the training phase in determining the overall carbon footprint. This shift indicates that the focus and methodology of AI environmental impact assessment are undergoing a fundamental transformation.

To address these emerging challenges and the growing need for consistent evaluation frameworks, this paper organizes the concepts, boundaries, and methodologies of AI carbon footprint assessment. For this scoping review, an extensive search and synthesis were conducted across academic and technical sources published up to October 31, 2025, focusing on studies, reports, and open-source tools related to the carbon footprint assessment of AI systems. Searches were performed in multidisciplinary databases such as Google Scholar, Web of Science, Scopus, and arXiv, and were complemented by a review of technical documentation for widely used Python-based carbon accounting libraries including CodeCarbon, CarbonTracker, and MLCO2 Impact. Grey literature and policy reports from organizations such as the IEA, ITU, and MIT Technology Review were also included to capture the most recent developments.

The collected materials were not selected on a systematic basis, and no claim is made that the evidence reviewed is exhaustive. Rather, the goal was to map the breadth of methods, metrics, and tools used to quantify AI-related emissions across model training, inference, and infrastructure operations. The findings were categorized and synthesized around the following analytical themes: (i) standardization and classification of AI carbon measurement methodologies and tools, (ii) comparative analysis of environmental impacts in the AI training and inference phases, and (iii) critical challenges, transparency requirements, and future governance directions.

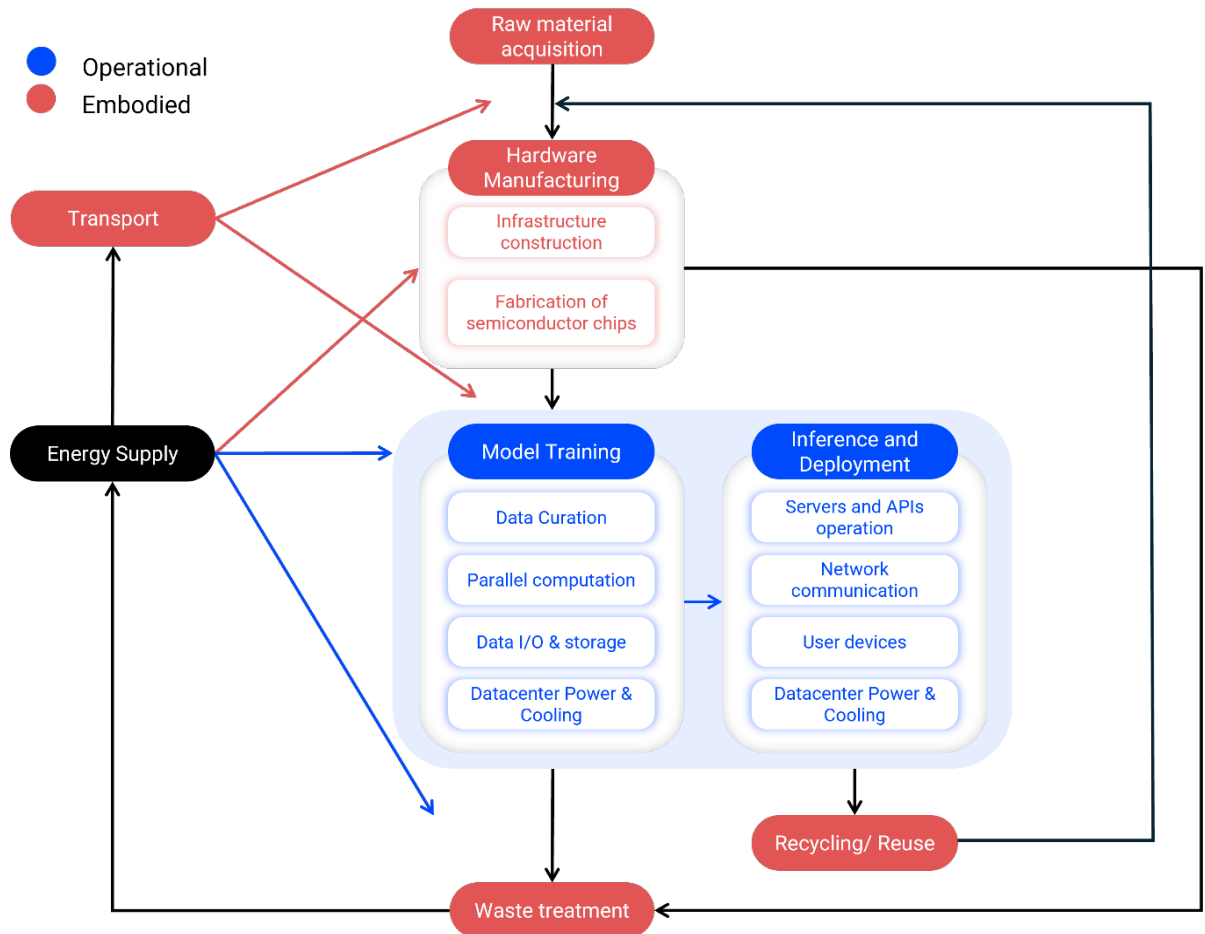

**Figure 1. Life Cycle of AI system (Conceptual Framework adapted from ISO 14040:2006)**

## 2. Carbon Footprint Assessment of AI

### 2.1. The Need for Standardization in AI Carbon Footprint Measurement

Carbon footprint assessment shares substantial conceptual similarities with LCA, and carbon footprint can be regarded as a simplified form of LCA that focuses solely on a single impact category, namely climate change (Clabeaux et al., 2020; Pattara et al., 2012) (see Figure 2). Both approaches aim to quantify the environmental impacts associated with a product, process, or system across its entire life cycle. However, because most carbon footprint studies limit their analysis to greenhouse gas emissions expressed in $CO_2$-equivalent values generated during the training or inference stages of a single AI model, they have been criticized for failing to fully capture broader environmental repercussions and higher-order impacts (Pattara et al., 2012; Plociennik et al., 2025; Wenmackers, 2024). In contrast, LCA encompasses a much wider range of environmental indicators, including water use, land occupation, resource depletion, and toxicity. Despite these differences, carbon footprint assessment should not be viewed merely as a reduced metric but rather as a foundational and practical entry point for integrating AI systems into the broader LCA framework. Accordingly, carbon footprint measurement for AI systems holds significant importance, as it contributes to greenhouse gas management in the short term while providing a pathway toward more comprehensive and multidimensional environmental impact assessments in the long term.

Current approaches to measuring the carbon footprint of AI remain uncertain and fragmented. Unlike traditional industrial processes, AI systems possess distinctive and complex characteristics. For example, the energy use boundaries for model training and inference are not clearly defined (ITU, 2025), and the resulting estimates vary widely depending on hardware architecture, methods for measuring electricity consumption, algorithmic implementation, and operational practices. In addition, studies rely on different indicators such as GPU usage, electricity consumption in kilowatt-hours, or $CO_2$-equivalent emissions, which makes comparison difficult. Regional variations in power grid mixes further exacerbate uncertainty. Moreover, when internal data from cloud service providers and companies are not disclosed, it becomes difficult to verify or reproduce reported estimates (A. S. Luccioni et al., 2025). As a result, the current measurements are often not comparable and have limited reliability.

For these reasons, standardization is essential. Establishing a common framework would improve the comparability of studies, strengthen accountability across academia and industry, and enable policymakers to design effective governance mechanisms (Wenmackers, 2024). Furthermore, standardization extends beyond technical convenience and can serve a normative role by embedding environmental responsibility into the development and deployment of AI. Consequently, the carbon footprint of AI should not be viewed merely as a technical metric but rather as a core component of sustainable AI governance.

The need for such efforts is increasingly recognized at the international level. The Greenhouse Gas Protocol provides general principles for carbon accounting, yet it still lacks guidance tailored specifically to digital and AI systems. The International Organization for Standardization (ISO) recently published the TR 20226:2025 document, which explicitly incorporates sustainability considerations for AI by addressing evaluation indicators related to the full life cycle of AI systems, including computational load, resource use, carbon impacts, pollution, disposal, and operational characteristics. In the European Union, policymakers have also emphasized the importance of tracking the environmental impacts of AI systems, and supplementary legislative provisions are being explored to regulate sustainability dimensions (Kroet, 2024). Organizations such as the Green Software Foundation are additionally developing methodologies to quantify and reduce the carbon intensity of software operations[1]. Together, these developments illustrate the urgency of establishing transparent and harmonized evaluation frameworks for AI, a technology that is rapidly proliferating across sectors.

As a starting point for advancing such standardization efforts, this review synthesizes diverse methodologies proposed in existing studies, identifies areas of inconsistency, and highlights the need for integration. By doing so, it seeks to bridge fragmented measurement approaches and contribute to building a foundation for incorporating AI into comprehensive life cycle–based sustainability assessment frameworks over the long term.

---

[1] *Green Software Foundation*, https://sci.greensoftware.foundation/, Accessed Oct 02. 2025

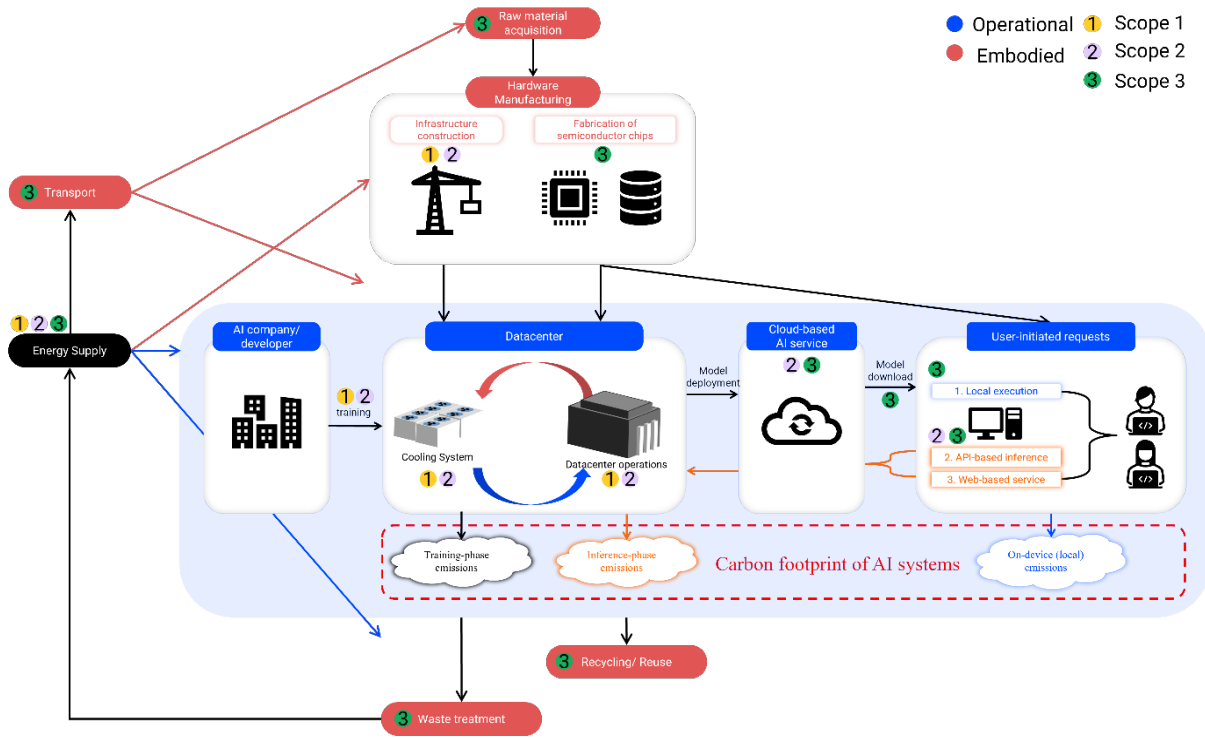

**Figure 2. Life Cycle and Carbon Footprint of AI System (Conceptual Framework adapted from ISO 14040:2006)**

## 2.2. Calculating the Carbon Footprint of AI

The operational carbon footprint of AI is generally calculated by multiplying electricity consumption (energy) by the carbon intensity of electricity generation. Carbon intensity refers to the amount of CO₂ emitted per unit of electricity produced (kWh) and varies by region and time depending on the energy mix of the power grid, including the proportions of coal, natural gas, and renewable energy. Therefore, even when the same amount of electricity is consumed, the resulting carbon footprint may differ depending on where and when the electricity is used.

$$Carbon\ Footprint = Energy \times Carbon\ Intensity$$

Where:

**Carbon Footprint**: The amount of carbon dioxide ($CO_2$) emissions generated by AI.

**Energy**: The actual amount of electricity (kWh) consumed by AI.

**Carbon Intensity**: The amount of $CO_2$ emissions (kg$CO_2$/kWh) produced per 1 kWh of electricity consumed. This value varies depending on the country, region, and energy mix (proportions of coal, gas, renewable energy, etc.).

Electricity consumption during the operational phase, which includes model training and inference, is commonly approximated as Energy ≈ Power × Runtime × PUE. Here, Power refers to the average electrical power (W) consumed by IT equipment, Runtime (h) is the actual duration of training or inference, and PUE (Power Usage Effectiveness) is a data center efficiency metric that incorporates facility-level overhead such as cooling, power conversion, and distribution losses. Applying PUE is essential because it reflects not only the electricity consumed by servers but also the total facility-wide energy demand, making it a realistic and widely used method for operational carbon estimation.

$$Energy = Power \times Run\,time \times PUE$$

Where:

**Power**: The power consumption (W) of hardware devices such as GPU, CPU, and DRAM.

**Runtime**: The total time (h) required for model training or inference.

**PUE**: A data center efficiency index that reflects not only the electricity consumed by the servers themselves but also cooling and power loss. (A value closer to 1 indicates higher efficiency)

The measured value of Power is highly dependent on the boundaries and instrumentation methods employed (Argerich & Patiño-Martínez, 2024; Elsworth et al., 2025). While some studies account only for GPU power consumption, the contributions of other key components such as CPUs, DRAM, storage, and networking equipment cannot be ignored. In particular, CPU and memory usage become more dominant during data preprocessing, model loading, and input-output operations. The choice of power measurement approach is another major source of variation. At the software and driver layers, NVML or nvidia-smi provides GPU power readings, while Intel RAPL offers estimates for CPU and DRAM consumption. These tools allow component-level breakdowns but may miss certain loads. At higher layers of measurement, server-level power can be monitored using Baseboard Management Controllers (BMCs), rack-level power through Power Distribution Units (PDUs) (Jay et al., 2023), and facility-level electricity consumption via building main meters. Although these upper-layer methods achieve higher total accuracy, they make it difficult to attribute energy use to specific components.

The reporting perspective for training and inference should also be distinguished. In the training stage, emissions can vary significantly depending on whether only the electricity consumption of a single training run is measured or whether the system boundary includes the total cost of auxiliary processes such as hyperparameter search, neural architecture search, failed runs, checkpointing, and data preprocessing (Patterson et al., 2021; Thompson et al., 2021). In the inference stage, if N denotes the total number of requests within the measurement interval, the carbon footprint per request can be defined as (Energy × Carbon intensity)/N. This value is highly sensitive to batch size, QPS (queries per second), latency targets, and resource utilization (Elsworth et al., 2025; Jegham et al., 2025; Samsi et al., 2023). For product or policy comparisons, one possible approach is to amortize training emissions across the total anticipated number of inference requests over the model's entire life cycle (that is, distributing the one-time training emissions across all future inference requests) and then combine this with operational emissions to obtain the total carbon footprint per request.

Ultimately, although carbon footprint estimation may appear straightforward as a simple multiplication, the actual results are determined by boundary definitions, measurement practices, and data choices. To ensure reproducibility and comparability, it is essential to clearly document the resource components included, the measurement layers and intervals, the data sources and temporal resolution of PUE and carbon intensity, and the specific elements included in the system boundaries for training and inference. When possible, uncertainties or confidence intervals should be reported, and operational emissions should be transparently separated from embodied emissions (Liu & Zhai, 2025).

## 3. Classification of Tools by Measurement Method

The measurement of the carbon footprint of AI originally began with estimating the energy consumption of machine learning training processes (Bieser, 2024; Morand et al., 2024). In earlier systems, most energy use was concentrated in the training stage. However, as the use of generative AI has expanded rapidly, the importance of measuring emissions during inference has increased substantially. Existing measurement methodologies were primarily developed to assess the training phase of machine learning or deep learning models, yet recent approaches have been designed specifically for inference. An important point is that measurement results differ depending on the operating system, hardware specifications, and whether the workload is executed on a personal computer or on a server. No single method is universally correct, and the choice of measurement approach should be determined by the goals and constraints of researchers and policymakers.

Current methods for measuring the carbon footprint of AI models can be broadly categorized into three groups. These include estimation approaches based on model specifications, direct hardware-based measurement methods, and sensor-assisted approaches that rely on operating system or software instrumentation.

Model-based estimation is the simplest approach and relies on predefined models to predict energy consumption. Tools such as Green Algorithms and ML CO2 Impact estimate energy use and carbon emissions based on user-provided information such as CPU and GPU specifications and model training time. Although this does not reflect actual power use precisely, it is useful for obtaining a rough magnitude of expected consumption. These models are effective for illustrating potential scale, but their results are highly sensitive to underlying assumptions and can become inaccurate due to accumulated estimation errors.

Physical instrumentation provides the most reliable measurements because it captures the actual power consumed by hardware devices. External power meters include tools such as Omegawatt and Raritan PDUs, which measure real-time power flow by inserting a device between the power line or leveraging built-in monitoring features of server-rack PDUs. These instruments are typically used to measure the power consumption of an entire server node. Internal measurement devices include Baseboard Management Controllers (BMCs) and tools such as PowerMon2. These devices are attached directly to the motherboard or specific components, enabling finer-grained measurement of CPU, memory, motherboard, and other subsystem-level power consumption.

Sensor-based measurement refers to approaches that rely on sensor data provided by the operating system or underlying software layers, which in turn collect information from sensors embedded within the hardware. For CPU and DRAM sensors, tools such as Intel RAPL (Linux), Intel Power Gadget (Windows), and powermetrics (macOS) are commonly used. Modern CPUs integrate performance monitoring counters (PMCs) and fully integrated voltage regulators (FIVRs), allowing power estimation to be supplemented with direct measurements. For GPUs, tools such as nvidia-smi, NVML, and pynvml provided by NVIDIA are widely used. These tools retrieve sensor data embedded within the GPU to report metrics such as power draw, temperature, and clock frequency. In data center environments, more specialized tools such as DCGM are employed to manage and monitor multiple GPUs at scale. For system-wide sensor access, tools like ioreg (macOS) and powerstat (Linux) aggregate data collected by battery controllers, power-management chips, and related components to provide an operating-system-level summary. Although these sensor-based approaches have improved overall measurement granularity, substantial methodological inconsistencies remain across studies, limiting comparability. Table 1 shows the characteristics of these measurement methods.

**Table 1. Comparison of Measurement Methodologies for AI System Energy Consumption and Carbon Emissions**

| Category | Model-based estimation | Sensor-based measurement | Physical instrumentation |
|---|---|---|---|
| Measurement Principle | Estimates energy consumption and carbon emissions using input parameters such as CPU/GPU specifications, RAM, and training duration | Collects sensor data from the operating system or hardware-embedded sensors (CPU, GPU, system-level sensors) through software | Directly measures actual power consumption using external power meters or internal current sensors |
| Tools | Green Algorithms, ML CO2 Impact, EcoLogits | Intel RAPL, Nvidia-smi, powermetrics, CodeCarbon, CarbonTracker, Experiment Impact Tracker | Omegawatt, Raritan PDU, BMC, PowerMon2 |
| Accuracy | Sensitive to assumptions and subject to cumulative estimation errors | Varies depending on sensor precision and depends on OS and driver implementation | Highest accuracy due to direct measurement |
| Data Collection Granularity | Model-level (estimated from CPU/GPU utilization) | Component-level (per-CPU/GPU power, temperature, clock frequency, etc.) | Server-node or component-level (CPU/GPU/memory, etc.) |
| Required Equipment | None (software- or web-based inputs) | CPUs/GPUs with embedded sensors; OS-level access privileges | External power meters, server-integrated meters, BMC, etc. |
| Cost and Accessibility | Low cost and highly accessible | Moderate (no extra hardware but setup required) | High cost and requires controlled lab or data center environment |
| Temporal Resolution | Low (averaged over entire training run) | High (real-time, second-level measurements) | Very high (detailed real-time power profiling) |
| Strengths | Simple, fast, suitable for large-scale comparisons | Easy to deploy, integrated with OS-level data | Most reliable, suitable for fine-grained hardware analysis |
| Limitations | Large estimation errors; dependent on assumptions | Dependent on OS and drivers; difficult to compare across studies | Requires specialized equipment and controlled environments |
| Suitable Use Cases | Policy-scale estimations, high-level academic comparison | Real-time monitoring, academic benchmarking | Precision experiments, hardware efficiency evaluation |

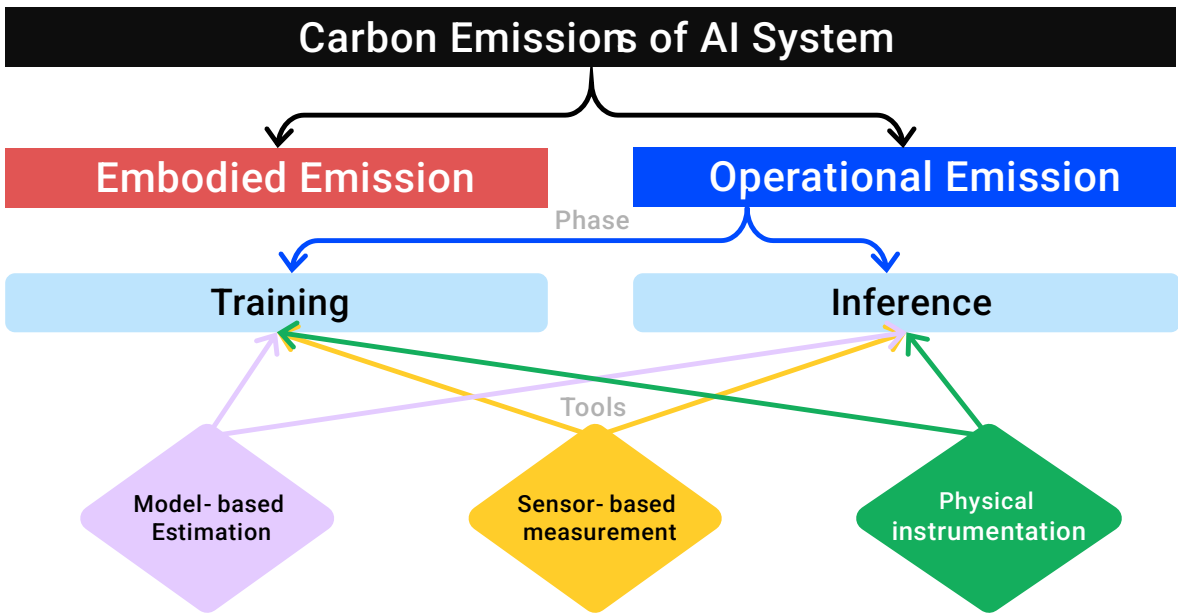

**Figure 3. Carbon Emissions of Generative AI and measuring tools**

## 4. Training phase

The training phase has long been recognized as the most energy-intensive component in the environmental impact of AI systems (IEA, 2025). Since the early days of machine learning, numerous efforts have been made to improve training efficiency, reduce resource consumption, and mitigate computational complexity (Verdecchia et al., 2023). These efforts have led to the development of a wide range of measurement and estimation tools for quantifying and managing carbon emissions. As large-scale model training has become increasingly common, quantitatively assessing the carbon footprint of the training phase has emerged as a key starting point for discussions on AI sustainability.

### 4.1. Tools for estimating the training phase of AI

Current tools for assessing the carbon footprint of the training phase can be categorized into three groups: (1) model-based estimation, (2) sensor-based measurement, and (3) hybrid approaches.

#### 4.1.1. Tools: Model-based estimation

This approach estimates carbon emissions using minimal input information. It typically uses regression equations, empirical models, or machine learning-based prediction formulas, offering high accessibility and simplicity.

- ML Emissions Calculator (Lacoste et al., 2019): A web-based tool that calculates $CO_2$-equivalent emissions by incorporating regional grid emission factors based on inputs such as server location, GPU type, training duration, data center PUE, and renewable energy certificates (RECs). It allows quick estimation of the environmental impact of large-scale model training with minimal inputs.
- Cumulator (Trébaol, 2020): An open-source Python package that estimates energy use from computation and data transfer during machine learning training and converts it into carbon emissions. It visualizes the performance-to-carbon trade-off by displaying carbon cost alongside model accuracy. Since it relies largely on runtime-based estimation, it functions more as a decision-support tool than a precise measurement method.

- Green Algorithms (Lannelongue et al., 2021): A web interface that computes carbon emissions associated with CPU, GPU, and cloud workloads using input values such as runtime, hardware specifications, and network configuration. As one of the representative model-based estimation tools, it provides a convenient way to assess carbon footprints across various computing environments.
- LLMCarbon (Faiz et al., 2023): An end-to-end estimation tool that covers the training, inference, and storage stages of LLMs. It calculates operational emissions by combining FLOPs, hardware efficiency, data center PUE, and grid carbon intensity, while also accounting for embodied carbon associated with the manufacturing of GPUs, CPUs, memory, and SSDs. By integrating the neural scaling law, it supports analyses of performance-carbon trade-offs.

Although this approach offers strong accessibility and simplicity, it may overestimate or underestimate actual emissions because it does not incorporate operational factors such as cooling, idle power, and parallelization effects.

### 4.1.2. Tools: Sensor-based measurement

Sensor-based measurement refers to methods that capture actual power consumption using hardware and software driver interfaces. For CPUs, Intel's RAPL (Running Average Power Limit) interface is commonly used to estimate power consumption, while for GPUs, NVIDIA's NVML (NVIDIA Management Library) API enables real-time collection of power usage data. Because these approaches rely on real measurements rather than estimates, they generally offer higher accuracy and reliability.

- Carbontracker (Anthony et al., 2020): Collects GPU, CPU, and DRAM power consumption using NVML and Intel RAPL and computes $CO_2$-equivalent emissions by incorporating regional grid carbon intensity and data center PUE. By monitoring only a subset of epochs, it can extrapolate total training emissions and provides carbon-saving interventions by recommending early stopping when thresholds are exceeded. This tool can be used across a variety of platforms, including clusters, desktop computers, and Google Colab notebooks.

- Energyusage (Lottick et al., 2019): Measures CPU power via RAPL and GPU power via the nvidia-smi CLI, and retrieves regional grid information through the GeoJS API to compute $CO_2$ emissions. Because it can evaluate not only training processes but also any arbitrary Python function, it is a highly general-purpose tool.
- Perun (Gutiérrez Hermosillo Muriedas et al., 2023): A Python library that uses RAPL, NVML, and psutil, while also integrating sensor data from the Lenovo XClarity Controller server management system. It measures power consumption from single machines to MPI-based multi-node systems and is well suited for large-scale parallel training analysis in high-performance computing environments.
- Experiment Impact Tracker (Henderson et al., 2020): Automatically tracks CPU, GPU, and DRAM power consumption with only a single line of additional code. It records process-level utilization and computes emissions by incorporating data center PUE and regional carbon intensity. It is applicable not only to AI training and inference but also to general-purpose code execution.
- Zeus (You et al., 2023): A tool that combines sensor-based measurement with power optimization capabilities. It measures GPU power via NVML while automatically exploring batch sizes and power limits to optimize energy efficiency in real time during deep neural network training. ML.Energy Leaderboard (ML.Energy Initiative, 2025) is a framework that uses Zeus to benchmark and compare GPU power usage during generative AI inference.
- pyJoules (Spirals Research Group, 2020): Measures CPU, GPU, and DRAM power consumption through RAPL and NVML by annotating sections of Python code with decorators. It enables fine-grained code-level power analysis, although it is limited in its ability to capture whole-system power consumption or differentiate between concurrent tasks.

Sensor-based measurement provides high precision because it reflects real power consumption. However, it faces access limitations in cloud and distributed environments. Additionally, most packages work only with Intel CPUs and NVIDIA GPUs, which restricts hardware accessibility depending on device support and driver compatibility.

### 4.1.3. Tools: Hybrid approach

Hybrid approaches combine model-based estimation with sensor-based measurement in order to compensate for the limitations of each method. For example, when a device does not provide access to Intel RAPL, power consumption may be estimated using TDP-based approximations, while GPU power may be partially corrected using sensor data retrieved from NVML. This approach reduces data-access constraints and can yield more realistic results than purely model-based estimation.

- CodeCarbon (Schmidt et al., 2021): Measures GPU power through NVML and CPU power through RAPL whenever possible, and falls back to TDP-based approximations when sensor access is unavailable. For RAM, the tool uses heuristic approximations. Total estimated energy consumption is combined with regional grid carbon intensity and data center PUE to compute final $CO_2$-equivalent emissions.
- Eco2AI (Budennyy et al., 2022): Measures GPU power using NVML and estimates CPU and RAM power consumption based on utilization, then multiplies the resulting energy estimate by PUE and carbon intensity to compute emissions. Unlike CodeCarbon, its estimation strategy differs when sensor-based measurements are unavailable. It also provides automated logging and real-time monitoring, offering practical estimations even in environments with limited measurement capabilities.

Hybrid methods strike a balance between precision and accessibility, functioning reliably in both cloud and local environments. Tools for measuring the carbon footprint of the training phase reflect the strengths of each paradigm: model-based approaches prioritize simplicity, sensor-based approaches prioritize empirical precision, and hybrid approaches serve as a practical compromise that accommodates real-world constraints.

### 4.2. Previous studies on Training phase

Early studies primarily relied on estimation-based methods to quantify environmental costs and raise awareness of the issue. Strubell et al. (2019) provided the first quantitative assessment of the carbon cost of NLP models such as Transformer, ELMo, BERT, and GPT-2. They measured GPU and CPU power using standard monitoring tools and estimated emissions based on typical data-center efficiency and grid carbon intensity. Their analysis showed that model size can lead to dramatic differences in emissions, with lightweight models producing relatively little carbon and large NAS-based models generating vastly more. Although Strubell et al. (2020) later revised the estimates downward by accounting for partial renewable-energy use on cloud platforms, the methodology still tended to overstate emissions due to limited alignment with real training environments.

Subsequent research began incorporating data center telemetry for empirical measurement. Patterson et al. (2021), using actual measurements from Google data centers, demonstrated that prior estimates were overestimated by up to a factor of 80. They showed that emissions could vary by several orders of magnitude depending on model design, hardware, data center efficiency, and grid characteristics. In an extended analysis, they highlighted that emissions for models like Evolved Transformer and Primer dropped sharply within two years, warning that predictions ignoring rapid technological progress are "fraught with peril" and emphasizing the need for ongoing measurement and reporting.

Parallel to these developments, new training paradigms and energy-efficiency comparisons also emerged. Qiu et al. (2023) offered the first quantitative analysis of energy use and emissions in federated learning, showing that communication can dominate overall emissions in some settings. This indicates that distributed learning is not inherently environmentally efficient and that its footprint depends heavily on data properties and local grid conditions..

Budennyy et al. (2022) used the eco2AI library to measure emissions from training the text-to-image models Malevich (1.3B) and Kandinsky (12B), finding more than a seventeen-fold difference in emissions depending on model size. They also showed that even changing the activation function (GELU versus few-bit GELU) could significantly affect training-time emissions. These examples illustrate how carbon-measurement tools can be useful for evaluating the environmental implications of model design choices.

Measurement frameworks were also expanded. Henderson et al. (2020) introduced Experiment Impact Tracker and built the "RL Energy Leaderboard" to compare reinforcement learning, image classification, and machine translation models. They found that emissions differed by up to thirtyfold depending on regional carbon intensity and that algorithm choices also mattered. Verma et al. (2024) showed that, despite differences in absolute estimates, many tools such as CarbonTracker and Eco2AI produced consistent emission trends across model families, highlighting the value of carbon-efficiency metrics for holistic sustainability evaluations.

Operational choices have also been shown to influence emissions. Ordoumpozanis and Papakostas (2024) conducted fine-tuning of Google's ViT model on medical imaging data and demonstrated that GPU oversizing and server location can significantly alter carbon footprints, emphasizing the importance of considering energy efficiency alongside performance. Wu et al. (2022), using telemetry from Facebook data centers, measured the full life-cycle emissions of language and recommendation models and found that inference accounted for 65 percent of total emissions, compared to 35 percent for training. They further showed that hardware–software co-optimization can reduce operational emissions by as much as a factor of 800.

In summary, research on the carbon footprint of the training phase has progressed from early estimation-based analyses to empirical validation, comparisons of training paradigms, and real-world model deployments. These studies show that emissions are shaped not only by model size but also by algorithmic choices, hardware, power grids, and operational practices. Nonetheless, the lack of standardized measurement frameworks, limited disclosure of corporate data, and the pace of technological change remain key challenges for establishing reliable and sustainable assessment systems.

## Table 2. Overview and Technical Specification of AI Carbon Footprint Measurement Tools

| Tool | ML CO2 Impact | Cumulator | Green Algorithms | LLMCarbon | CarbonTracker | Experiment Impact Tracker | Zeus | Energyusage | Perun | pyJoules | CodeCarbon | Eco2AI | EcoLogits | Energy Meter |
|---|---|---|---|---|---|---|---|---|---|---|---|---|---|---|
| **Development & Access** | | | | | | | | | | | | | | |
| Reference | (Lacoste et al., 2019) | (Trébaol, 2020) | (Lannelongue et al., 2021) | (Faiz et al., 2023) | (Anthony et al., 2020) | (Henderson et al., 2020) | (You et al., 2023) | (Lottick et al., 2019) | (Gutiérrez Hermosillo Muriedas et al., 2023) | (Spirals Research Group, 2020) | (Schmidt et al., 2021) | (Budennyy et al., 2022) | (Rincé & Banse, 2025) | (Argerich & Patiño-Martínez, 2024) |
| Tool Type | Online calculator | Embedded package (Python library) | Online calculator | Embedded package (Python library) | Embedded package (Python library) | Embedded package (Python library) | Embedded package (Python library) | Embedded package (Python library) | Embedded package (Python library) | Embedded package (Python library) | Embedded package (Python library) | Embedded package (Python library) | Embedded package (Python library) | Embedded package (Python library) |
| First (latest) release date | Aug. 2019 (Jul. 2022) | Jun. 2020 | Jul. 2020 (Jun. 2022) | Sep. 2023 (Jan. 2024) | Apr. 2020 (Jul. 2021) | Dec. 2019 (Jan. 2020) | Apr. 2023 | Dec. 2019 | Aug. 2023 | 2020 (2022) | Nov. 2020 (Sept. 2022) | 2022 | Jul. 2025 | Jun. 2024 |
| Embodied emission | no | no | no | Yes | no | no | no | no | no | no | no | no | Yes | no |
| **Environment & Scope** | | | | | | | | | | | | | | |
| Target Step | Training | Training | Training | Training, Inference, Experimentation, Storage | Training | Training, Inference | Training | Training | General Python function execution | General Python function execution | Training, Inference | Training | Inference | Inference |
| Target Models | DL | ML | Algorithms | LLM | DL | ML | DNN | General Python computation | DL | General Python computation | ML | ML | LLM | LLM |
| Scope | machine | Expanded Operation | process | LCA | machine | process | process | process | Expanded Operation | machine | machine/process | machine/process | LCA | process |
| OS compatibility | any | any | any | any | Linux (for RAPL access) | Linux (for RAPL access) / Mac OS | Linux (tested on CloudLab & Chameleon Cloud) | Linux (for RAPL access) | Linux (for RAPL access) | Linux (for RAPL access) | any | Linux, Windows, macOS | any | Linux (for RAPL/eBPF access) |
| Hardware compatibility | any | any | any | any | Intel RAPL, Nvidia NVML | Intel RAPL, Nvidia NVML | NVIDIA GPU | Intel RAPL, NVIDIA GPU | Intel RAPL, Nvidia NVML | Intel RAPL, Nvidia NVML | Intel/AMD (RAPL, Linux), NVIDIA GPU (NVML) | Intel/AMD CPU, NVIDIA GPU (NVML) | any | Intel RAPL, NVIDIA NVML |
| CPU Energy Estimation | - | TDP-based Hardware Modeling | TDP-based Hardware Modeling | - | RAPL files (Intel only) | RAPL / Power Gadget (Intel only) | - | Intel RAPL | Intel RAPL | Intel RAPL | RAPL / Power Gadget; 85 W if unknown | CPU TDP lookup via internal DB (3279 models); 100 W if unknown | Model-based estimation | RAPL (pyRAPL) |
| GPU Energy Estimation | TDP-based Hardware Modeling | TDP-based Hardware Modeling | TDP-based Hardware Modeling | TDP-based Hardware Modeling | pynvml | nvidia-smi | NVIDIA NVML | nvidia-smi CLI | NVIDIA NVML | pynvml, intel RAPL (iGPU) | pynvml | pynvml | Model-based estimation | NVML (pyNVML) |
| Memory Energy Estimation | - | - | 0.3725 W/GB of available memory (user-provided) | Average System Power-based Modeling (user-provided) | RAPL memory total (not per process) | RAPL or Power Gadget + psutil shared mem | - | - | Intel RAPL | RAPL DRAM domain (Intel only, not per process) | Heuristic based on RAM size/Number of RAM slots (v3) | 0.375 W/GB of used memory (via psutil) | Model-based estimation | RAPL (pyRAPL) |
| Communication Cost Estimation | - | 6.894 x 10^(-11) kWh/byte (1byte model, 2017) | - | - | - | - | - | - | psutil for network and filesystem I/O monitoring | - | - | - | - | Storage cost estimation (eBPF) |
| Default TDP | – / – | 250 / | 12 / 200 | – / – | – / – | – / – | – / – | – / – | – / – | – / – | 85 / – | 100 / – | – / – | – / – |

| | | | | | | | | | | | | | | | |
|---|---|---|---|---|---|---|---|---|---|---|---|---|---|---|---|
| **(CPU / GPU)** | | 250(NVIDIA GeForce GTX Titan X) | | | | | | | | | | | | | |
| **Usage Factor Handling** | Not computed | No (assumes 100%) | Yes (user-provided or assumes 100%) | Yes (19.7–47 % hardware efficiency by model) | Not computed | psutil for CPU/Mem, nvidia-smi for GPU | NVML for GPU | RAPL for CPU, nvidia-smi CLI for GPU | psutil for CPU/Mem, NVML for GPU load | Not computed | 50% assumed if default TDP used | psutil for CPU, pynvml for GPU | Not computed | Configurable (Idle removal) |
| **Coefficients** | | | | | | | | | | | | | | | |
| **PUE (Default / Configurable)** | N/A (Not used) | N/A (Not used) | 1.67 (default, configurable) | 1.09–1.12 (configurable) | 1.58 (default 2018, configurable) | 1.58 (default 2020, configurable) | N/A (Not used) | N/A (not used) | 1.05 (HoreKa system default, configurable) | N/A (not used) | Cloud only (default N/A) | 1.0 (default, configurable) | 1.2 (default, configurable) | - |
| **Emission Intensity (gCO₂eq/kWh)** | User-provided or cloud-specific | 447 (EU avg 2018) | 475 (World avg 2018) | region-specific(configurable) | 475 (2019 avg) | 301 (ElectricityMap zones Avg) | N/A (Not used) | region-specific(GeoJS APIs + EIA / eGRID 2016 regional data) | User-provided | N/A (not used) | region-specific(475, configurable) | region-specific(436.5, configurable) | Multi-criteria (GWP, ADP, PE, conficurable) | -- |

Note: Reorganized from (Bannour et al.; Bouza et al., 2023; Jay et al., 2023)

* Scope:

- LCA: This category includes tools that explicitly model or estimate the embodied carbon footprint arising from the manufacturing of computing hardware, in addition to the operational carbon footprint generated during AI model execution. These tools aim to provide an integrated assessment of environmental impacts across the full life cycle of AI systems, including training, inference, experimentation, and storage.

- Expanded operational: In addition to measuring the energy consumption of major computing hardware components such as CPUs, GPUs, and DRAM, these tools explicitly incorporate communication costs and data-transfer-related energy usage (e.g., WAN, I/O). This is typically achieved using specialized functions or external models, such as the "1-byte model," to account for network and data movement overhead.

- Process: This category includes tools that estimate the energy consumption attributed to a specific process or task by partitioning its share from the total machine-level consumption. Approaches often compute a usage factor (e.g., using psutil or nvidia-smi) or allocate consumption based on the memory exclusively used by the process.

- Machine: Tools in this category report the total energy consumption of an entire computing node obtained from hardware sensors (e.g., RAPL, NVML), or model whole-machine energy consumption using thermal design power (TDP) and execution time.

## 5. Inference phase

Inference refers to the stage in which an AI model is deployed and executed in real-world service environments. Although the energy consumed per task is lower than during training, the cumulative carbon footprint can be substantially larger due to long-term, large-scale repeated execution. With the widespread adoption of generative AI, the energy demand of inference has become increasingly important (Argerich & Patiño-Martínez, 2024). Moreover, while training can be scheduled during low-carbon hours or executed in low-emission regions, inference must be performed in real time. As a result, constraints related to latency, network bandwidth, privacy, and regulatory requirements limit opportunities for energy reduction (ITU, 2025).

The carbon footprint of the inference phase is fundamentally calculated using the same principle applied to the training phase, in which total energy consumption is multiplied by the carbon intensity of the local electricity grid. Because inference is repetitive and executed at massive scale, normalized metrics such as energy per inference or energy per token are commonly used.

1) Energy per inference

$$Energy\ per\ inference\ (Wh) = \frac{Total\ runtime\ energy\ (Wh)}{Inference\ count}$$

Where:

Total runtime energy (Wh): The total amount of energy consumed across all inference operations.
Inference count: The total number of inference operations executed under the same setting.

2) Energy per token

$$Energy\ per\ token\ (Wh) = \frac{Power\ \times\ Inference\ duration}{Number\ of\ tokens\ \times\ 3600}$$

Where:

Power (Wh): The power consumption of hardware devices (GPU, CPU, RAM, etc.) during inference.

Inference duration (s): The time required to perform a single inference.

Number of tokens: The total number of tokens generated during that inference.

**5.1. Tools for estimating the inference phase of AI**

There are relatively few tools developed exclusively for assessing the inference phase of generative AI. This is because most existing measurement tools are designed to cover both training and inference. For example, CodeCarbon, Experiment Impact Tracker, Perun, Zeus, EnergyUsage, and Eco2AI can track power consumption and emissions during inference using the same mechanisms employed for the training phase. However, tools specifically tailored to the characteristics of inference are beginning to appear, with EcoLogits and EnergyMeter being representative examples.

- EcoLogits (Rincé & Banse, 2025): An open-source Python library designed to quantify the environmental impacts of API-based inference for generative AI systems. It goes beyond energy use by estimating broader potential environmental burdens, including hardware manufacturing and data center operations. It provides per-request estimates of energy use, carbon emissions, global warming potential (GWP), abiotic depletion potential (ADP), and primary energy demand. EcoLogits adopts an LCA-based approach aligned with ISO 14044, estimating GPU cloud computation, cooling, storage, and other system components through a bottom-up modeling strategy. However, it uses model-based estimation instead of direct hardware measurement.
- Energy Meter (Argerich & Patiño-Martínez, 2024): An open-source tool for real-time measurement of CPU, GPU, memory, and storage power consumption during local LLM inference. It collects hardware-level measurements using Intel RAPL, NVIDIA NVML, and other sensor interfaces, enabling detailed comparisons of inference efficiency across different configurations such as model size, batch size, and quantization settings. Since it reports only energy values, users must apply regional carbon intensity separately to estimate $CO_2$-equivalent emissions.

EcoLogits represents a cloud-oriented, model-based estimation approach, while EnergyMeter reflects a sensor-based measurement approach. Together, these tools provide inference-specific frameworks that allow for more accurate and fine-grained evaluation of the environmental impacts associated with AI service deployment.

## 5.2. Benchmark-Based Comparative Studies

Interpreting inference efficiency requires clearly defined measurement boundaries and proper experimental control. Even for the same model, metrics normalized per prompt (or per token) can vary by several folds depending on whether the measurement boundary includes only the GPU or also encompasses CPU and RAM, or whether it extends further to end-to-end coverage, including idle machine energy and system-level overhead (see Figure 4). Recent studies show that end-to-end measurements can produce carbon-emission estimates up to 2.4 times higher than GPU-only measurements (Elsworth et al., 2025). As different tools and benchmarks adopt different measurement boundaries, comparing absolute values across studies becomes inherently challenging. However, because measurement boundaries and methodologies vary across studies, it is critical to interpret results with attention to each study's design assumptions rather than relying on the absolute values.

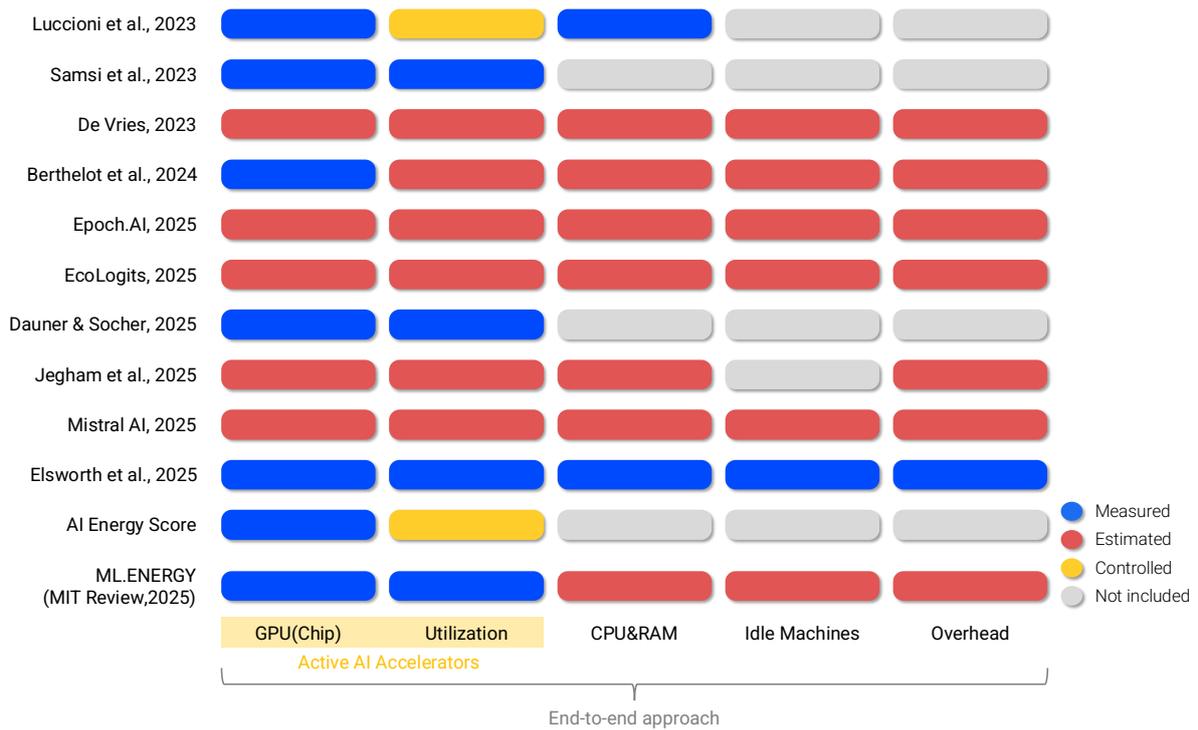

**Figure 4. Measurement Boundaries**

Recent benchmark studies can be grouped into four main approaches: (1) an indirect estimation (regression-based) approach that enables fast approximation across many models, (2) emphasizing standardized empirical measurement to maximize comparability, (3) attempts to approximate realistic operational behavior (steady-state simulation) by incorporating deployment constraints, (4) a new research direction presenting quantitative accuracy–energy trade-offs, highlighting the nonlinear relationship between performance and environmental impact.

EcoLogits represents a typical indirect estimation approach. It predicts GPU power consumption using regression models that incorporate variables such as accelerator type, CPU and DRAM specifications, and expected output length. Multiplying this estimate by average data center PUE yields the energy consumption per prompt. Although sensitive to hardware and workload assumptions and lacking direct measurement, estimation-based approaches provide a broader system-level view by accounting for data center overhead as well as GPU use. In contrast, Hugging Face's AI Energy Score adopts a standardized empirical approach. It runs text generation, image generation, and image captioning tasks with batch size 1 on controlled hardware configurations (NVIDIA H100 80GB), enforcing strict consistency in hardware, task type, and batch size (Luccioni et al., 2025). By providing periodically updated relative rankings across classes of models (grouped by required GPU type or count), this benchmark greatly enhances comparability. However, because it intentionally excludes high-efficiency batching and pipelining commonly used in real-world services, its scores may diverge from absolute values observed in production environments.

ML.Energy takes the opposite perspective, incorporating realistic batching and latency constraints to approximate deployment scenarios (ML.Energy Initiative, 2025). Using Zeus framework to measure GPU power in real time, it computes per-request energy across generative tasks. To bridge the gap between GPU only and end-to-end measurement, MIT Technology Review proposed a simple correction: normalize GPU power and apply a "double-scaling" factor of two, assuming that GPUs represent roughly half of total server power (O'Donnell & Crownhart, 2025). This adjustment implicitly covers colling and other system level loads and is served as a "pragmatic bridge" for interpreting boundary differences rather than an exact estimator. In short, while AI Energy Score prioritizes comparability, ML.Energy and MIT's adjustment method approximate absolute values based on realistic deployment scenarios.

Finally, Dauner and Socher (2025) evaluated both accuracy and emissions for fourteen LLMs under identical conditions (NVIDIA A100 80GB) using the 1,000-question Massive Multitask

Language Understanding dataset. Emissions ranged widely, from 27.7 gCO$_2$e per 1,000 prompts for Qwen-7B to 2,042.4 gCO$_2$e for DeepSeek-R1 70B, with generated token counts and reasoning complexity as major drivers. Moreover, they observed pronounced nonlinear trade-offs: models with higher accuracy exhibited sharply increasing energy consumption, demonstrating that performance–emissions relationships cannot be assumed to scale proportionally. By jointly reporting accuracy and emissions on a shared benchmark, this study reframes the definition of a "good" model as a multi-objective optimization problem involving both performance and environmental efficiency.

### 5.3. Determinants of Inference Carbon Footprint across Modalities and Prompts

**Text models**

In text generation models, model size is the most intuitive predictor of inference energy demand, since larger parameter counts generally require more computation for each generated token. This overall pattern is shown in Figure 5, where higher performing and larger models tend to cluster toward the upper left region, forming an apparent efficiency frontier: models with stronger performance typically lower energy efficiency. However, model size alone does not account for the full range of variation in inference emissions. Recent studies highlight additional determinants such as prompt length, reasoning complexity, and the number of output tokens, all of which significantly influence the energy cost of generating a single response.

MIT Technology Review and ML.Energy reported that LLaMA 3.1 8B model consumed approximately 57 joules per response (114 joules including cooling and non-GPU overhead), while the larger 405B model required about 6,700 joules. Yet these size differences capture only part of the picture. Jegham et al. (2025) showed that prompt length (for example, 10k tokens compared to 1.5k tokens) and reasoning depth can increase energy use by several orders of magnitude. Notably, even among models of comparable size, the difference can be extreme. GPT-4.1 nano required only 0.45 Wh under long-prompt conditions, whereas the reasoning-intensive DeepSeek-R1 consumed 33.6 Wh, despite belonging to a similar model class.

Similarly, Dauner and Socher (2025) found that the number of output tokens per question and the depth of reasoning are the strongest predictors of emissions across fourteen models tested under identical conditions. Their findings indicate that inference cost depends not only on the model architecture but also on how the model is prompted and how much reasoning it is required to perform.

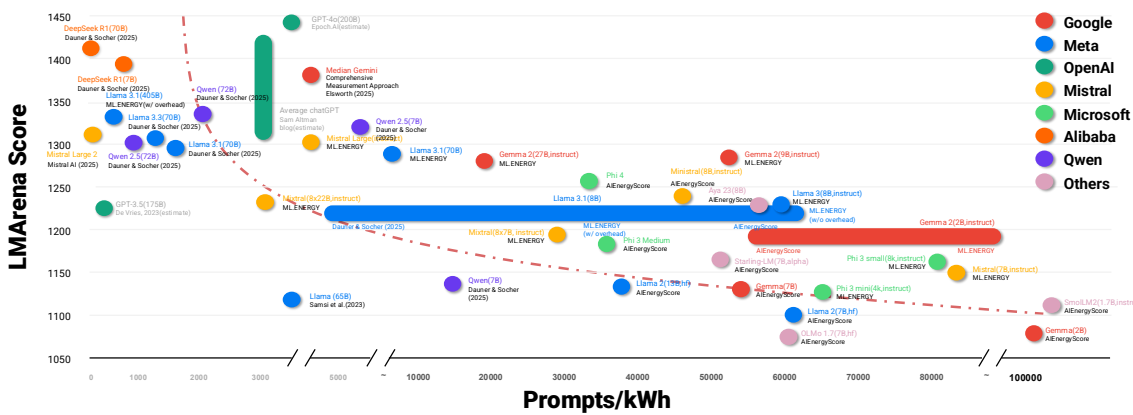

**Figure 5. Performance and energy efficiency landscape of LLMs** (Note: Values for LMArena score and prompts per kWh are synthesized from multiple empirical sources: (Dauner & Socher, 2025; De Vries, 2023; Elsworth et al., 2025; ML.Energy Initiative, 2025; Samsi et al., 2023; Sasha et al., 2025))

**Image and video models**

When modalities transition from text to images or video, the dimensionality and computational requirements increase substantially. As a result, the baseline energy demand per request becomes considerably higher.

According to MIT Technology Review, Stable Diffusion 3 Medium requires approximately

1,141 joules of GPU energy to produce a 1024×1024 image (roughly 2 billion parameters) (O'Donnell & Crownhart, 2025). Berthelot et al. (2025) extended the analysis by evaluating image generation as an end-to-end service using LCA approach. Their integrated system boundary, which included user devices, networks, hosting infrastructure, and data management, resulted in an estimated footprint of 360 tCO$_2$eq and about 7.8 gCO$_2$eq per request. Taken together, these studies show that single inference energy values capture only the computational core of image generation, whereas service level assessments highlight the substantial additional impacts arising from infrastructure and usage patterns.

Video generation is even more computationally intensive. Experiments by Luccioni on CogVideoX showed that generating a single five second clip required about 3.4 megajoules of GPU energy (O'Donnell & Crownhart, 2025), which is approximately seven hundred times higher than what is needed to generate a high quality image. This contrast illustrates how the carbon cost per request can increase nonlinearly as modalities become more complex.

**Interaction with serving environment and boundary**

Prompt complexity and modality interact with the serving environment, and this interaction greatly shapes reported emissions. Inference efficiency can also vary depending on serving design, including batching strategies and latency constraints that affect how computation is scheduled in practice.

Deployment environment plays a critical role as well (Masciari & Napolitano, 2025). Li et al. (2025) showed that edge inference on the Samsung S24 (Snapdragon 8 Gen 3 NPU) reduced energy use by more than 90 percent compared to Google Colab (A100 40GB) across text, image, and OCR tasks. For large commercial systems, data center efficiency factors such as PUE,

WUE, CIF, and renewable energy procurement exert a stronger influence on absolute emissions, therefore Jegham et al. (2025) incorporated these coefficients into a standardized assessment of commercial inference. Elsworth et al. (2025) reported 0.24 Wh per prompt for Google Gemini under a full-stack boundary that included GPU, CPU, DRAM, idle capacity, and cooling. Under a traditional GPU-only boundary, the value dropped to 0.10 Wh, underscoring how strongly results depend on boundary definition.

Overall, the inference carbon footprint is determined by (see Table 3): (i) the computational scale associated with each modality (text < image < video), (ii) prompt structure, including token length and reasoning depth, (iii) serving design, such as batching, latency constraints, and scheduling, (iv) deployment environment and infrastructure efficiency (for example, cloud versus edge), and (v) boundary definitions, including whether measurements focus narrowly on accelerators or encompass full-stack energy use.

**Table 3. Summary of Factors Influencing the Environmental Impact of AI Model Inference**

| Category | Key Factor | Description | Related Studies | Direct of Impact |
|---|---|---|---|---|
| Modality | Text < Image < Video | Increasing data and computing requirements raise energy, carbon, and water use per request | (O'Donnell & Crownhart, 2025) | Nonlinear increase in energy use |
| Prompt Complexity | Token length, reasoning depth | Long prompts, chain-of-thought, and multi-step reasoning increase the number of decoding operations and reasoning computations | (Dauner & Socher, 2025; Jegham et al., 2025) | Exponential increase in energy use |
| Operational Design | Batch size, Latency constraints, pipelining | Efficient batching and scheduling can reduce energy consumption per request | (ML.Energy Initiative, 2025; O'Donnell & Crownhart, 2025; Sasha et al., 2025) | Reduction in per-request energy |
| Serving Environment | Cloud versus Edge, infrastructure | Edge reduced network and data center overhead; data center efficiency and | (Jegham et al., 2025; Masciari & Napolitano, 2025) | Reduction relative to cloud inference |

| | efficiency (PUE, WUE, CIF) | renewable procurement determine absolute emissions | | |
|---|---|---|---|---|
| Boundary Definition | GPU-only vs end-to-end | Inclusion of CPU, DRAM, cooling, networking, and idle capacity more than two-fold difference | (Elsworth et al., 2025) | Large variation in reported emissions |

# 6. Discussion

## 6.1. Methodological Limitations and Technology Bias

Current studies on AI carbon footprint measurement continue to face challenges in methodological clarity and reproducibility. At the corporate level, major companies estimate emissions based on internal telemetry data from CPUs, GPUs, RAM, and facility-level electricity use. However, these data are rarely disclosed or standardized, which makes it difficult for researchers to reproduce results under equivalent conditions. In contrast, individual researchers often rely on small-scale GPUs or rented cloud resources, conducting simplified measurements that differ widely in baseline assumptions and procedural choices. As a result, measurements across studies remain difficult to compare.

Variation in measurement approach also contributes to substantial inconsistencies. Different approaches use different methods, resulting in varying levels of precision and comparability. Measurements can be taken at the process, machine/node, or life-cycle level, and the chosen boundary significantly affects the reported emissions. Many studies do not clearly specify which subsystems are included, such as whether the boundary accounts for only GPU power or also CPU, RAM, idle power, and system overhead. The absence of a standardized measurement protocol remains a fundamental barrier to transparency and comparability across studies.

A further concern is the strong dependence on specific hardware ecosystems. Most existing work relies heavily on NVIDIA GPUs, and widely used measurement tools depend on driver interfaces such as nvidia-smi or NVML. While this reflects current market realities, it also means that measurement practices are effectively tied to the technological ecosystem of a single vendor. Such dependence risks reinforcing technology-specific bias and may compromise the generalizability of carbon accounting methods as hardware becomes more diverse. These challenges highlight the need for harmonized and hardware-agnostic methodologies that can reduce dependence on specific proprietary telemetry channels.

**6.2. Misalignment with System Realities and the Absence of Social Context**

A second limitation is the misalignment between current research practices and real-world AI system operation. Most studies continue to concentrate on the training phase, whereas the environmental impact of inference remains comparatively understudied. This is problematic because LLMs are used repeatedly after deployment, and cumulative emissions from inference can exceed those from training. Despite this, empirical inference measurement tools and benchmark datasets remain scarce, and existing studies often limit their scope to model-level efficiency comparisons.

Differences in serving environments also play an important role in shaping emissions. Inference can occur in local edge environments or cloud data centers, and these settings differ substantially in overhead, energy mix, and infrastructure efficiency. Most academic studies rely on local GPU measurements, while users of commercial AI services predominantly rely on cloud-based API calls. This gap leads to inconsistencies between experimental conditions and actual usage patterns. Future work should therefore compare inference across local, cloud, and API-based settings, and empirically assess how each operational mode influences emissions.

Although model efficiency has improved significantly in recent years and per-request emissions have decreased accordingly, such improvements do not necessarily translate into reductions in aggregate energy demand. Increased efficiency may stimulate higher usage, giving rise to Jevons' paradox and rebound effects (S. Luccioni et al., 2025b). To date, no empirical research has examined how user behavior, adoption patterns, or societal demand interact with AI system efficiency to determine overall environmental outcomes. This lack of social and behavioral analysis represents a major gap in the current literature and signals that technological efficiency alone cannot guarantee reductions in total emissions.

**6.3. Sustainability of Evaluation Frameworks**

Finally, while interest in applying LCA to AI systems has grown (Berthelot et al., 2024; Luccioni et al., 2023; Plociennik et al., 2025), current practices remain partial and fragmented. Unlike traditional products or services, AI systems are composed of interconnected layers of data pipelines, models, hardware components, and cloud infrastructure. These systems evolve rapidly, and their interdependencies make it difficult for traditional LCA approaches to capture the full environmental impact. Most existing studies focus on training or inference as isolated subsystems and fall short of evaluating cradle-to-gate or full life cycle impacts. Even within operational emissions, end-to-end systemwide telemetry remains limited, and research integrating operational and embodied emissions is still at an early stage.

A further difficulty is that technological progress is advancing faster than the development of assessment frameworks. The emergence of generative and multimodal AI systems has increased architectural complexity, yet current methodologies largely rely on static, model-level evaluations. Such approaches do not adequately reflect dynamic factors such as algorithmic improvements, changes in data scale, or evolving infrastructure configurations. As a result, evaluation systems may underestimate the environmental impact of rapidly changing AI ecosystems.

# 7. Conclusion

This scoping review has catalogued the measurement tools and frameworks currently available to researchers and practitioners. Drawing on the issues identified in the preceding sections, future research on AI carbon footprint and LCA needs to advance along four major directions.

First, standardizing methodologies and enhancing transparency. One of the primary reasons for the difficulty in comparing results across studies is the heterogeneity of measurement boundaries, measurement methods, and operating conditions. Future work should establish a Generative AI Environmental Footprint Reporting Checklist aligned with ongoing international standardization efforts under ISO/IEC and ITU. Such a checklist may include: (1) task and modality specification, (2) prompt and output token counts, (3) hardware specifications, (4) operational conditions, (5) measurement tools and techniques used, (6) definition of system boundaries, and (7) units for reporting results.

To mitigate dependence on specific corporate technologies and reduce data bias, a vendor-neutral carbon accounting framework capable of spanning heterogeneous hardware environments must also be developed.

More than 84% of commercialized large models released since 2022 provide no information on energy or carbon emissions. Such data opacity undermines comparability and verifiability, and risks distorting policy decisions and technological development pathways. Transparent disclosure and data sharing by companies are therefore essential. With accumulated empirical measurements and open datasets, AI inference can evolve into a service that balances performance, environmental impact, and cost in a sustainable manner.

Second, measuring designs that incorporate behavioral and operational factors. Most existing studies remain limited to static comparisons of model performance and per-request emissions. However, future measurement designs need to reflect dynamic factors such as user behavior and real-world workload mixtures. Traffic patterns, prompt length and complexity, latency requirements, and prompt types should be quantitatively examined for their causal impact on per-request energy consumption and emissions. Beyond prompt-based or model-based measurement, higher-order effects arising from usage behavior and social diffusion also need to be addressed (Wenmackers, 2024). A comprehensive demand-based analysis that considers user prompt habits, model choices, expectations of output quality, and usage frequency can provide a basis for policy interventions and carbon mitigation strategies. These factors are also essential for evaluating the efficiency of model and hardware selection, as well as service-level deployment strategies such as batching, caching, and routing.

Third, end-to-end observability and full LCA across AI services. A more accurate understanding of AI's carbon footprint requires end-to-end observability within operational emissions. Measurement boundaries should expand beyond the model, CPU, and GPU to include the entire service stack, covering networking, storage, caching, cooling, power delivery, orchestration layers, and idle resources. In addition, an end-to-end monitoring framework is needed to integrate energy flows and carbon exchanges across the computing continuum, including edge, fog, and cloud environments (Li et al., 2025; Masciari & Napolitano, 2025; Patterson et al., 2024). A LCA framework that incorporates both embodied emissions and operational emissions is also required (Liu & Zhai, 2025). Current research on AI carbon footprints largely focuses on the operational phase, which overlooks environmental burdens from semiconductor fabrication, data center construction, and network infrastructure. Even if model efficiency improves, total environmental impact may increase due to new GPU production and large-scale data center expansion. LCA therefore provides a necessary foundation for verifying real net environmental benefits and burdens. Traditional LCA approaches, however, rely on static system boundaries and therefore fail to capture dynamic factors such as rapid architectural advances, improvements in model efficiency, or evolving data-center infrastructure. Future work should develop an adaptive LCA framework that incorporates real-time changes in model architecture, dataset scale, hardware generations, and system configurations. Such a framework would allow more flexible and sustainable assessments of AI's environmental impacts.

Finally, beyond carbon emissions, multidimensional sustainability metrics, including energy consumption, water usage, resource depletion, embedded material impacts, and electronic waste, need to be considered. AI systems consume not only electricity but also water (for cooling), minerals, and chemical resources. An AI service with low carbon intensity may still impose significant water burdens if its cooling requirements are high. This highlights the need for integrated, multi-dimensional environmental impact assessment frameworks that incorporate carbon–water–materials perspectives. In addition, evaluation systems should jointly report model accuracy and emissions to support balanced decision-making based on Pareto optimality (Dauner & Socher, 2025; Jegham et al., 2025). In the short term, metrics such as Energy Efficiency Score (model performance per unit energy consumed) can help quantify trade-offs between performance and environmental costs. In the long term, normative metric systems may be required to guide model development toward reducing environmental burdens even at the expense of modest performance sacrifices.

**References**

Anthony, L. F. W., Kanding, B., & Selvan, R. (2020). Carbontracker: Tracking and Predicting the Carbon Footprint of Training Deep Learning Models. https://doi.org/10.48550/arxiv.2007.03051

Argerich, M. F., & Patiño-Martínez, M. (2024). Measuring and Improving the Energy Efficiency of Large Language Models Inference. *IEEE Access*, *12*, 80194-80207. https://doi.org/10.1109/access.2024.3409745

Bannour, N., Ghannay, S., Névéol, A., & Ligozat, A.-L. (2021, 2021). Evaluating the carbon footprint of NLP methods: a survey and analysis of existing tools.

Berthelot, A., Caron, E., Jay, M., & Lefèvre, L. (2024). Estimating the environmental impact of Generative-AI services using an LCA-based methodology. *Procedia CIRP*, *122*, 707-712.

Berthelot, A., Caron, E., Jay, M., & Lefèvre, L. (2025). Understanding the Environmental Impact of Generative AI Services. *Communications of the ACM*, *68*(7), 46-53. https://doi.org/10.1145/3725984

Bieser, J. (2024, 2024). AI and Climate Protection: Research Gaps and Needs to Align Machine Learning with Greenhouse Gas Reductions.

Bouza, L., Bugeau, A., & Lannelongue, L. (2023). How to estimate carbon footprint when training deep learning models? A guide and review. *Environmental Research Communications*, *5*(11), 115014. https://doi.org/10.1088/2515-7620/acf81b

Budennyy, S. A., Lazarev, V. D., Zakharenko, N. N., Korovin, A. N., Plosskaya, O. A., Dimitrov, D. V., Akhripkin, V. S., Pavlov, I. V., Oseledets, I. V., Barsola, I. S., Egorov, I. V., Kosterina, A. A., & Zhukov, L. E. (2022). eco2AI: Carbon Emissions Tracking of Machine Learning Models as the First Step Towards Sustainable AI. *Doklady Mathematics*, *106*(S1), S118-S128. https://doi.org/10.1134/s1064562422060230

Clabeaux, R., Carbajales-Dale, M., Ladner, D., & Walker, T. (2020). Assessing the carbon footprint of a university campus using a life cycle assessment approach. *Journal of Cleaner Production*, *273*, 122600.

Dauner, M., & Socher, G. (2025). Energy costs of communicating with AI. *Frontiers in Communication*, *10*. https://doi.org/10.3389/fcomm.2025.1572947

De Vries, A. (2023). The growing energy footprint of artificial intelligence. *Joule*, *7*(10),

2191-2194. https://doi.org/10.1016/j.joule.2023.09.004

Elsworth, C., Huang, K., Patterson, D., Schneider, I., Sedivy, R., Goodman, S., Townsend, B., Ranganathan, P., Dean, J., Vahdat, A., Gomes, B., & Manyika, J. (2025). Measuring the environmental impact of delivering AI at Google Scale. https://doi.org/10.48550/arxiv.2508.15734

Faiz, A., Kaneda, S., Wang, R., Osi, R., Sharma, P., Chen, F., & Jiang, L. (2023). LLMCARBON: MODELING THE END-TO-END CARBON FOOTPRINT OF LARGE LANGUAGE MODELS. https://doi.org/10.48550/arxiv.2309.14393

Gaur, L., Afaq, A., Arora, G. K., & Khan, N. (2023). Artificial intelligence for carbon emissions using system of systems theory. *Ecological Informatics*, *76*, 102165.

Gutiérrez Hermosillo Muriedas, J. P., Flügel, K., Debus, C., Obermaier, H., Streit, A., & Götz, M. (2023). perun: Benchmarking Energy Consumption of High-Performance Computing Applications. In (pp. 17-31). Springer Nature Switzerland. https://doi.org/10.1007/978-3-031-39698-4_2

Henderson, P., Hu, J., Romoff, J., Brunskill, E., Jurafsky, D., & Pineau, J. (2020). Towards the systematic reporting of the energy and carbon footprints of machine learning. *Journal of machine learning research*, *21*(248), 1-43.

IEA. (2025). *Energy and AI*.

ITU. (2025). *Measuring what matters: How to assess AI's environmental impact*.

Jay, M., Ostapenco, V., Lefevre, L., Trystram, D., Orgerie, A.-C., & Fichel, B. (2023, 2023). An experimental comparison of software-based power meters: focus on CPU and GPU.

Jegham, N., Abdelatti, M., Elmoubarki, L., & Hendawi, A. (2025). How Hungry is AI? Benchmarking Energy, Water, andCarbon Footprint of LLM Inference. https://doi.org/10.48550/arxiv.2505.09598

Kroet, C. (2024). EU Commission aims to regulate environmental impact of AI through delegated act. *Euronews*. https://www.euronews.com/next/2024/11/28/eu-commission-aims-to-regulate-environmental-impact-of-ai-through-delegated-act?utm_source=chatgpt.com

Kurisu, K. (2015). Application of Life Cycle Assessment (LCA) to Assess Actual Environmental Burdens Driven by PEBs. In *Pro-environmental Behaviors* (pp. 99-

129). https://doi.org/10.1007/978-4-431-55834-7_5

Lacoste, A., Luccioni, A. S., Schmidt, V., & Damdres, T. (2019). Quantifying the Carbon Emissions of Machine Learning. https://doi.org/10.48550/arxiv.1910.09700

Lannelongue, L., Grealey, J., & Inouye, M. (2021). Green Algorithms: Quantifying the Carbon Footprint of Computation. *Advanced Science*, *8*(12), 2100707. https://doi.org/10.1002/advs.202100707

Li, P., Islam, M. J., & Ren, S. (2025). A Case Study of Environmental Footprints for Generative AI Inference: Cloud versus Edge. *ACM SIGMETRICS Performance Evaluation Review*, *53*(2), 21-26. https://doi.org/10.1145/3764944.3764950

Liu, H., & Zhai, J. (2025). Carbon Emission Modeling for High-Performance Computing-Based AI in New Power Systems with Large-Scale Renewable Energy Integration. *Processes*, *13*(2), 595. https://doi.org/10.3390/pr13020595

Lottick, K., Susai, S., Friedler, S. A., & Wilson, J. P. (2019). Energy Usage Reports: Environmental awareness as part of algorithmic accountability. https://doi.org/10.48550/arxiv.1911.08354

Luccioni, Gamazaychikov, Strubell, Hooker, Jernite, Wu, & Mitchell. (2025). *AI Energy Score Leaderboard*. Hugging Face. Retrieved Nov 05 from https://huggingface.co/AIEnergyScore

Luccioni, A. S., Gamazaychikov, B., da Costa, T. A., & Strubell, E. (2025). Misinformation by Omission:The Need for More Environmental Transparency in AI. https://doi.org/10.48550/arxiv.2506.15572

Luccioni, S., Strubell, E., & Crawford, K. (2025a). From efficiency gains to rebound effects: The problem of jevons' paradox in AI's polarized environmental debate. Proceedings of the 2025 ACM Conference on Fairness, Accountability, and Transparency,

Luccioni, S., Strubell, E., & Crawford, K. (2025b, 2025). From Efficiency Gains to Rebound Effects: The Problem of Jevons' Paradox in AI's Polarized Environmental Debate.

Luccioni, S., Viguier, S., & Ligozat, A.-L. (2023). Estimating the carbon footprint of bloom, a 176b parameter language model. *Journal of Machine Learning Research*, *24*(253), 1-15.

Masciari, E., & Napolitano, E. V. (2025). Environmental Sustainability of AI: Estimating CO2e Emissions Across Cloud, Edge, and Fog Paradigms. In (pp. 409-418). Springer

Nature Singapore. https://doi.org/10.1007/978-981-96-1483-7_33

ML.Energy Initiative. (2025). *ML.Energy Leaderboard*. Retrieved 10.15 from https://ml.energy/leaderboard/

Morand, C. e., Ligozat, A.-L., & ́eol1, A. e. N. e. (2024). How Green Can AI Be? A Study of Trends in Machine Learning Environmental Impacts. https://doi.org/10.48550/arxiv.2412.17376

O'Donnell, J., & Crownhart, C. (2025). We did the math on AI's energy footprint. Here's the story you haven't heard. *MIT Technology Review.* https://www. techn ology review. com/2025/05/20/1116*, 3*, 27.

Ordoumpozanis, K., & Papakostas, G. A. (2024). Green AI: Assessing the Carbon Footprint of Fine-Tuning Pre-Trained Deep Learning Models in Medical Imaging. 2024 International Conference on Innovation and Intelligence for Informatics, Computing, and Technologies (3ICT),

Patel, P., Choukse, E., Zhang, C., Goiri, Í., Warrier, B., Mahalingam, N., & Bianchini, R. (2024, 2024). Characterizing Power Management Opportunities for LLMs in the Cloud.

Pattara, C., Raggi, A., & Cichelli, A. (2012). Life Cycle Assessment and Carbon Footprint in the Wine Supply-Chain. *Environmental Management*, *49*(6), 1247-1258. https://doi.org/10.1007/s00267-012-9844-3

Patterson, D., Gilbert, J. M., Gruteser, M., Robles, E., Sekar, K., Wei, Y., & Zhu, T. (2024). Energy and Emissions of Machine Learning on Smartphones vs. the Cloud. *Communications of the ACM*, *67*(2), 86-97. https://doi.org/10.1145/3624719

Patterson, D., Gonzalez, J., Le, Q., Liang, C., Munguia, L.-M., Rothchild, D., So, D., Texier, M., & Dean, J. (2021). Carbon emissions and large neural network training.

Plociennik, C., Watjanatepin, P., Van Acker, K., & Ruskowski, M. (2025). Life Cycle Assessment of Artificial Intelligence Applications: Research Gaps and Opportunities. *Procedia CIRP*, *135*, 924-929.

Qiu, X., Parcollet, T., Fernandez-Marques, J., Gusmao, P. P., Gao, Y., Beutel, D. J., Topal, T., Mathur, A., & Lane, N. D. (2023). A first look into the carbon footprint of federated learning. *Journal of machine learning research*, *24*(129), 1-23.

Rasheed, M. Q., Yuhuan, Z., Haseeb, A., Ahmed, Z., & Saud, S. (2024). Asymmetric

relationship between competitive industrial performance, renewable energy, industrialization, and carbon footprint: Does artificial intelligence matter for environmental sustainability? *Applied Energy*, *367*, 123346.

Rincé, S., & Banse, A. (2025). EcoLogits: Evaluating the Environmental Impacts of Generative AI. *Journal of Open Source Software*, *10*(111), 7471. https://doi.org/10.21105/joss.07471

Samsi, S., Zhao, D., McDonald, J., Li, B., Michaleas, A., Jones, M., Bergeron, W., Kepner, J., Tiwari, D., & Gadepally, V. (2023, 2023). From Words to Watts: Benchmarking the Energy Costs of Large Language Model Inference.

Sasha, L., Gamazaychikov, B., Strubell, E., Hooker, S., Jernite, Y., Wu, C.-J., & Mitchell, M. (2025). *AI Energy Score Leaderboard*. Hugging Face. Retrieved 11.05 from https://huggingface.co/AIEnergyScore

Schmidt, V., Goyal, K., Joshi, A., Feld, B., Conell, L., Laskaris, N., Blank, D., Wilson, J., Friedler, S., & Luccioni, S. (2021). CodeCarbon: estimate and track carbon emissions from machine learning computing. *Cited on*, *20*.

Schneider, I., Xu, H., Benecke, S., Patterson, D., Huang, K., Ranganathan, P., & Elsworth, C. (2025). Life-Cycle Emissions of AI Hardware: A Cradle-To-Grave Approach and Generational Trends. https://doi.org/10.48550/arxiv.2502.01671

Shehabi, A., Hubbard, A., Newkirk, A., Lei, N., Siddik, M. A. B., Holecek, B., & Sartor, D. (2024). *2024 United States Data Center Energy Usage Report*. https://dx.doi.org/10.71468/p1wc7q

Spirals Research Group. (2020). *pyJoules*. In (Version 0.5.1) https://pypi.org/project/pyJoules/

Strubell, E., Ganesh, A., & McCallum, A. (2019, 2019). Energy and Policy Considerations for Deep Learning in NLP.

Strubell, E., Ganesh, A., & McCallum, A. (2020). Energy and Policy Considerations for Modern Deep Learning Research. *Proceedings of the AAAI Conference on Artificial Intelligence*, *34*(09), 13693-13696. https://doi.org/10.1609/aaai.v34i09.7123

Thompson, N. C., Greenewald, K., Lee, K., & Manso, G. F. (2021). Deep Learning's Diminishing Returns: The Cost of Improvement is Becoming Unsustainable. *IEEE Spectrum*, *58*(10), 50-55. https://doi.org/10.1109/mspec.2021.9563954

Trébaol, T. (2020). *CUMULATOR — a tool to quantify and report the carbon footprint of machine learning computations and communication in academia and healthcare*.

Verdecchia, R., Sallou, J., & Cruz, L. (2023). A systematic review of Green AI. *WIREs Data Mining and Knowledge Discovery*, *13*(4). https://doi.org/10.1002/widm.1507

Verma, A., Kumar Singh, S., Kumar Sah, R., Misra, R., & Singh, T. N. (2024, 2024). Performance Comparison of Deep Learning Models for CO2 Prediction: Analyzing Carbon Footprint with Advanced Trackers.

Wang, Q., Li, Y., & Li, R. (2024). Ecological footprints, carbon emissions, and energy transitions: the impact of artificial intelligence (AI). *Humanities and Social Sciences Communications*, *11*(1). https://doi.org/10.1057/s41599-024-03520-5

Wenmackers, S. (2024). Assessing the Ecological Impact of AI. https://doi.org/10.48550/arxiv.2507.21102

Wu, C.-J., Raghavendra, R., Gupta, U., Acun, B., Ardalani, N., Maeng, K., Chang, G., Aga, F., Huang, J., & Bai, C. (2022). Sustainable ai: Environmental implications, challenges and opportunities. *Proceedings of machine learning and systems*, *4*, 795-813.

You, J., Chung, J.-W., & Chowdhury, M. (2023). Zeus: Understanding and optimizing {GPU} energy consumption of {DNN} training. 20th USENIX Symposium on Networked Systems Design and Implementation (NSDI 23),